The latex file and postscript(ps) file of our preprint are between the begin
and end comments, respectively.
 Could you please strip the ps file off carefully ?

****** begin of LaTex file **********************************************
\documentstyle [12pt]{article}

\def\thalf{{\textstyle{\frac{1}{2}}}}
\def\tquar{{\textstyle{\frac{1}{4}}}}
\def\pj{\hspace{-0.27cm}}
\begin{document}
\begin{titlepage}
\pagestyle{empty}
\begin{center}
\begin{large}
{{\bf Approximate Treatment of Hermitian Effective Interactions and
a Bound on the Error}}
\end{large}
\vskip 1cm
Ryoji Okamoto$^{a}$, Kenji Suzuki$^{a}$, P.J. Ellis$^{b}$,
Jifa Hao$^{c}$, Zibang Li$^{c}$ and T.T.S. Kuo$^{c}$\\
\vskip 1cm
\begin{em}{\small
$^a$ Department of Physics, Kyushu Institute of Technology, Kitakyushu 804 }\\
\end{em}
\begin{em}{\small
$^b$School of Physics and Astronomy, University of Minnesota, Minneapolis, MN
55455, USA}\\
\end{em}
\begin{em}{\small
$^c$Department of Physics, State University of New York at Stony Brook,
Stony Brook, NY 11794, USA}\\
\end{em}
\vskip 0.7cm
{\bf Abstract}
\end{center}
The Hermitian effective interaction can be well-approximated
by $(R+R^{\dagger})/2$ if the eigenvalues of $\omega^{\dagger}\omega$
are small or state-independent (degenerate), where $R$ is the standard
non-Hermitian effective interaction and $\omega$ maps the
model-space states onto the excluded space.  An error bound on
this approximation is given.
\vskip0.7cm
\vskip-18.5cm
\end{titlepage}

Much effort has been made to calculate the shell-model effective interactions
in nuclei from a realistic nucleon-nucleon interaction.  In spite of a great
deal of progress[1-5] in this field of
 physics, attention has been directed almost entirely to the well-known
non-Hermitian form, which we label $R$ here.
However, the empirical or phenomenological shell-model effective
interactions have been assumed to be Hermitian.  Therefore
direct comparison between the theoretical and empirical effective
interactions might cause confusion. The formal theory of constructing
 a Hermitian effective interaction, which we denote here by $W$,
 has been developed since des Cloizeaux\cite{des} and Brandow's\cite{bhb}
original works.
  Recently an improved approach was introduced\cite{suz} and was applied
to the calculation of Hermitian effective interactions, starting with modern
meson-exchange nucleon-nucleon interactions, by several authors\cite{keh}.  In
their study it has been observed that the
non-Hermiticity was rather small and $(R+R^{\dagger})/2$, referred to as
$W_{app}$, was a very good approximation to the exact Hermitian $W$.
This  raises the general question as to under what conditions the
approximation $W\simeq W_{app}$ might be reliable. The origin of and bounds
on the non-Hermiticity of $R$ has already been discussed by us\cite{soek}.
 The main  purpose of
 this note is to derive an explicit relation with which the validity of the
approximation $W\simeq W_{app}$ can be qualitatively examined. An application
of our relations to a model matrix problem will be made, and we shall discuss
the results.

We define two projection operators $P$ and $Q$ according to
the usual definitions {\it i.e.}, $P$ and $Q$ project a state onto the model
space and its complement,
respectively, and they satisfy $P+Q=1$. Let $d$ be the dimension of the $P$
space.  We write
$d$ of the true eigenstates of $H$ to be reproduced from the $P$-space
effective interaction as\cite{sl}
\begin{equation}
|\Phi_k\rangle\equiv(P+Q)|\Phi_k\rangle=|\phi_k\rangle+
\omega|\phi_k\rangle\;,
\end{equation}
where $|\phi_{k}\rangle$ is the $P$-space component of $|\Phi_{k}\rangle$.
The operator $\omega$ maps the $P$-space state
$|\phi_k \rangle$ onto the $Q$ space..
The operator $\omega$ is related to the usual wave operator $\Omega$ as
$\omega=\Omega-P$.
The operator $\omega$ is written explicitly as
\begin{equation}
\omega=\sum\limits_k Q|\Phi_k\rangle\langle\tilde{\phi}_k|P\;.
\end{equation}
Here $|\tilde{\phi}_k\rangle$ is the biorthogonal complement to the model
 space wave function $|\phi_k\rangle$, {\it i.e.},
$\langle\tilde{\phi}_k|\phi_i\rangle=\delta_{ki}$.
The Hamiltonian $H$ is divided into two parts, the unperturbed part $H_0$
and the perturbation $V$.   Using $\omega$
the non-Hermitian effective interaction $R$ can be written as
\begin{equation}
R=PVP+PVQ\omega\
\end{equation}
which is equivalent to the usual definition of the non-Hermitian effective
interaction $PV\Omega$.

 The Hermitian effective interaction
 $W$ may be written in the $|\alpha \rangle$ basis as\cite{suz,keh}
\begin{equation}
\langle \alpha|W|\beta\rangle=D(\alpha,\beta)
\left\{\sqrt{\mu^2_{\alpha}+1}\langle\alpha|R|\beta \rangle +
\sqrt{\mu^2_{\beta}+1}\langle \alpha|R^{\dagger}|\beta \rangle\right\}\;,
\end{equation}
where $|\alpha \rangle$ ($|\beta \rangle$) and $\mu_{\alpha}
(\mu_{\beta})$ are
given through the eigenvalue equation of $\omega^{\dagger}\omega$
\begin{equation}
\omega^{\dagger}\omega|\alpha \rangle = \mu^2_{\alpha}|\alpha \rangle\;,
\end{equation}
and
\begin{equation}
D(\alpha,\beta)=\left\{\sqrt{\mu^2_{\alpha}+1}+
\sqrt{\mu^2_{\beta}+1}
\right\}^{-1}\;.
\end{equation}
{}From the definition of $\omega$ in Eq.(2) we easily see that the operator
$\omega^{\dagger}\omega$ is a Hermitian operator acting in the $P$ space and it
 has positive or zero eigenvalues.

The Hermitian form of $W$ in Eq.(4) is formally exact.  However, it has been
known that in some cases $W$ can be well approximated by $W_{app}$\cite{keh}.
  In order to
measure the deviation of $W_{app}$ from the exact $W$, we introduce a quantity
\begin{equation}
\Delta W= \sum\limits_{ij}|\langle i|W-W_{app}|j \rangle|\ ,\ {\rm where}\ \
W_{app}=\thalf(R+R^{\dagger})\;,
\end{equation}
and $|i\rangle$, $|j\rangle$ and $|k\rangle$ are the
basis states, which are the eigenstates of the
 unperturbed Hamiltonian $H_0$.
Using a relation in the $|\alpha\rangle$ basis
\begin{equation}
\frac{\sqrt{\mu^2_{\alpha}+1}}{\sqrt{\mu^2_{\beta}+1}}\langle\alpha
 |H_0+R|\beta\rangle=\frac{\sqrt{\mu^2_{\beta}+1}}
{\sqrt{\mu^2_{\alpha}+1}}
\langle\alpha |H_0+R^{\dagger}|\beta\rangle \;,
\end{equation}
where both sides are equal to $\langle \alpha|H_0+W|\beta \rangle$\cite{keh},
the deviation $\Delta W$ is converted to
\begin{eqnarray}
\Delta W&\pj=&\pj\tquar\sum\limits_{ij}\bigg|\sum\limits_{\alpha\beta}
C(\alpha,\beta)^2
\nonumber\\
&&\times\left\{\frac{\langle i|\alpha \rangle\langle \beta|j \rangle}
{\mu^2_{\beta}+1}\langle \alpha|H_0+R| \beta \rangle +
\frac{\langle i|\beta \rangle\langle \alpha |j \rangle}
{\mu^2_{\alpha}+1}\langle \alpha|H_0+
R^{\dagger}|\beta \rangle \right\}\bigg|\;,
\end{eqnarray}
where
\begin{equation}
C(\alpha,\beta)=\sqrt{\mu^2_{\alpha}+1}-\sqrt{\mu^2_{\beta}+1}\;.
\end{equation}
In general, the matrix element of effective Hamiltonian $H_0+R$ is bounded,
that is,
\begin{equation}
|\langle \alpha|H_0+R| \beta \rangle|=|\langle \alpha|H|\beta \rangle+
\mu_{\beta}\langle \alpha|V|\nu_{\beta} \rangle|\leq(1+\mu_{\beta})V_0,
\end{equation}
where $V_0$ is the maximum
value of the matrix element of $PHP$ and $PVQ$, and
\begin{equation}
|\nu_{\beta} \rangle=\mu_{\beta}^{-1}\omega|\beta \rangle\;.
\end{equation}
{}From Eqs.(9) and (11) and since $|\langle i|\alpha \rangle \langle \beta
|j \rangle|\leq 1$, it can be  proved that there exists
 a constant $W_0$ such that
\begin{equation}
\Delta W \leq W_0 Z_w,
\end{equation}
where
\begin{equation}
Z_w =\tquar\sum\limits_{\alpha\beta}C(\alpha,\beta)^2
\left\{\frac{\mu_{\alpha}+1}{\mu^2_{\alpha}+1} +
\frac{\mu_{\beta}+1}{\mu^2_{\beta}+1}\right\}\;.
\end{equation}
Here the constant $W_0$ is independent of $\mu_{\alpha}$ and has a
bound given by $W_0 \leq d^2 V_0$ (we recall that $d$ is the dimension of the
$P$ space).
{}From Eqs.(13) and (14) we may say that the magnitude of the
deviation $\Delta W$ is determined by the eigenvalues, $\mu_{\alpha}^2$, of
 $\omega^{\dagger}\omega$.  If they are small then
$C(\alpha,\beta)$ is small and therefore
$Z_w$ and $\Delta W$ are also small.  This may be understandable
naturally because, when the $\mu_{\alpha}$ are small, the matrix elements
$\langle \alpha |PVQ\omega |\beta \rangle$ are small and the effective
interactions $R$, $R^{\dagger}$ and $W$ are almost the same.
Eq.(14) gives us another criterion, namely if the $\mu_{\alpha}$ are
state-independent, that is, the $\mu_{\alpha}$ are close to a constant,
then $C(\alpha,\beta)$ is again small and hence so is $\Delta W$.
Similar criteria govern the degree of non-Hermiticity of
 the non-Hermitian effective interaction $R$\cite{soek}.

As has been discussed in Ref.\cite{soek}, a  set of states
\begin{equation}
|\zeta_{\alpha} \rangle=\frac{|\alpha \rangle+
\mu_{\alpha}|\nu_{\alpha}
 \rangle}{\sqrt{\mu^2_{\alpha}+1}}\;, (\alpha=1,2,...,d)
\end{equation}
span a $d$-dimensional orthogonal subspace, denoted by $S$, where
$|\alpha \rangle$ are the eigenstates of $\omega^{\dagger}\omega$ and
$|\nu_{\alpha} \rangle$
are the $Q$-space states defined in Eq.(11).
One can prove that the space $S$ is an invariant subspace with respect to $H$.
In other words, the diagonalization of $H$ within $S$ yields $d$ true
eigenstates $|\Phi_k \rangle$ of $H$ in Eq.(1).  Since $|\Phi_k \rangle$ can
be expressed in terms of only $|\zeta_{\alpha} \rangle$, the ratio
$\mu_{\alpha}$
of mixing of $|\nu_{\alpha} \rangle$ into $|\alpha \rangle$ is preserved in
any true eigenstate of $H$.
The validity of the approximation $W\simeq W_{app}$ is related to the state
dependence of the mixing ratio $\mu_{\alpha}$ between the $P$- and $Q$-space
states.  If the $Q$-space states mix into the $P$-space states with
 a constant mixing ratio,
the approximation $W\simeq W_{app}$ can be acceptable.

For a simple illustration of the analysis of the approximation
 $W\simeq W_{app}$ we consider a model problem with an exactly solvable
Hamiltonian.  The model Hamiltonian we shall use is $H=H_0+V$, where the
unperturbed part is $H_0={\rm diag}(1,1,3,9)$ and the perturbation is
$V=[V_{ij}]$ given by
$$V=\left(\matrix{0&5x&-5x&5x\cr5x&25x&5x&
-8x\cr-5x&5x&-5x&x\cr5x&-8x&x&-5x\cr}\right)\;,$$
where $x$ is a parameter that we shall vary. This Hamiltonian
was introduced originally by Hoffmann {\it et al.}\cite{model},
but the matrix elements $V_{13}=V_{31}$ and $V_{24}=V_{42}$ are
changed from their original values of zero.  The Hamiltonian $H$ has
 already been applied to the study
 of non-Hermiticity in the effective interaction by the authors\cite{soek}.

 We shall take the lowest two eigenstates of $H_0$
to be our model space ($P$ space). In principle, provided the
true eigenstates have non-zero components in the $P$ space, any set of the true
 eigenvalues of $H$ can be reproduced
from the $P$ space effective Hamiltonian, {\it i.e.}, $H_0+R$ with the
non-Hermitian
effective interaction $R$ or $H_0+W$ with the Hermitian form $W$. However,
in the present study we discuss only the effective interactions
which reproduce the true eigenstates with the largest $P$-space overlaps,
because our main aim is not to show the variety of effective
interactions but to clarify the validity of the approximation
 $W\simeq W_{app}$.

The effective interaction $R$  reproducing the largest $P$-space overlaps
can be calculated according to the iteration scheme of Krenciglowa and
Kuo\cite{emt}, which corresponds to the resummation of folded
diagrams\cite{klr}
to infinite order.  In the next step by applying Eq.(4), we obtain the
Hermitian effective interaction $W$.  In Eq.(4) $W$ is given in the
$|\alpha \rangle$ basis, but it will be easy to rewrite it in terms of the
$|i \rangle$ basis states which are eigenfunctions of $H_0$. Then $\Delta W$
can be computed exactly and compared with the bound of Eqs.(13) and (14).

The exact solution for the eigenstates of $H$ shows that for small
$x$ the states with largest $P$-space overlaps are the first and
second lowest states.  As $x$ increases, the largest $P$-space overlap states
 change to the eigenstates with the first and third lowest eigenvalues for
0.0689$<x<$0.2655, while for $x>$0.2655 they are the first and fourth
eigenvalues.  In Table 1 the exact eigenvalues $E_k$ to be reproduced are
shown for six values of $x$ together with the $P$-space
overlaps $I_k$ defined by
\begin{equation}
I_k=\frac{\sum_{i=1}^2\langle i|\Phi_k \rangle^2}
{\langle\Phi_k|\Phi_k\rangle}\;,
\end{equation}
where $|\Phi_k \rangle$ is the true eigenstates of $H$.  The approximate
Hermitian effective interaction $W_{app}$ and the exact one
$W_{exact}$(=$W$), as well as the deviation $\Delta W$ of
$W_{app}$ from $W_{exact}$, are presented.
The quantity $\delta E$ represents a measure of the deviation of the
eigenvalues $E^{app}_k$ of $H_0+W_{app}$ from the exact values $E_k$,
which is defined by
\begin{equation}
\delta E= \sum_{k=1}^2 |E^{app}_k-E_k|\;.
\end{equation}
The values of $\mu^2_{\alpha}$, the eigenvalues of
 $\omega^{\dagger}\omega$ in Eq.(5), are also presented.

For $x$ of 0.04 the overlap of the true eigenstates with the model
space is close to 1 and the matrix elements of $W_{app}$ are identical
to $W_{exact}$
to the accuracy quoted.  Correspondingly $\Delta W$ is very small, as is
$\delta E$.  As $x$ increases, the overlaps $I_k$ become smaller and
conversely the deviation $\Delta W$ starts to grow.  However, at $x$
 of 0.22 the deviation $\Delta W$, as well as $\delta E$, is reduced
drastically, although the overlaps $I_k$ are not so large.  At this value of
$x$ we see that the eigenvalues $\mu_{\alpha}$ are almost the same.
Therefore we may say that if the $\mu_{\alpha}$ are nearly degenerate,
$\Delta W$ becomes small and the approximation $W\simeq W_{app}$
is justified.  This fact is an expected result of the theoretical
prediction for $\Delta W$ in Eqs.(13)--(14). For the larger values of $x$ in
Table 1 we notice that, even though we reproduce the eigenstates with the
largest $P$-space overlap, the deviation $\Delta W$ and the error in the
approximate eigenvalues $\delta E$ become quite sizeable. Our discussion
indicates that this is due to one of the $\mu_{\alpha}$ being large and the
other small.

In order to see that $\Delta W$ is bounded by $W_0Z_w$, as
shown in Eq.(13), we show $\Delta W$ and $W_0Z_w$ as functions of $x$
in Fig.1.  We here take $W_0$ to be $0.6+15x$ which is 3/20 of
$d^2V_0$.  In Fig.1 there appear two ``level-crossing" points,
{\it i.e.}, $x=0.0689$ and $x=0.2655$.   At these points the
second eigenstate with the predominant $P$-space component moves
from the second lowest state to the third and subsequently
from the third to the fourth. The curves are discontinuous at these
``level-crossing" points.  Fig.1 shows clearly that $\Delta W$
never exceeds $W_0Z_w$, that is, $\Delta W$ is bounded by
$W_0Z_w$.  Since $W_0$ is a constant when $x$ is fixed,
the validity of the approximation $W\simeq W_{app}$ is governed
by $Z_w$ which is a function of $\mu_{\alpha}$.  The characteristics of the
function $Z_w$ are that if all $\mu_{\alpha}$ are small,
$Z_w$ becomes small, and if the $\mu_{\alpha}$ are state independent,
{\it i.e.}, close to a constant, $Z_w$ can also be small.  If
$\mu_{\alpha}$ are large and have strong state dependence,
$Z_w$ becomes large and the approximation $W\simeq W_{app}$
will be poor.  This situation is quite similar to the case of
the degree of non-Hermiticity in the effective interaction
$R$ as has been discussed in Ref.\cite{soek}.  In general, we may
say that if the degree of non-Hermiticity is small, the approximation
 $W\simeq W_{app}$ will be good.

In summary, the accuracy of approximating the Hermitian effective interaction
 $W$ by $W_{app}=(R+R^{\dagger})/2$ with the usual non-Hermitian
 effective interaction $R$ is best judged by the eigenvalues $\mu^2$
of $\omega^{\dagger} \omega$, where $\omega$ is the
operator which maps the model-space states onto the excluded
space.  Both the theoretical prediction and a model calculation show
 that if the eigenvalues $\mu^2$ are small or state-independent, the
approximation is
justified.  Our study shows that the accuracy of the approximation
$W \simeq W_{app}$ cannot be judged merely by the magnitude of
the $Q$-space overlaps in the true eigenstates to be reproduced.

\vskip.5cm
This work was initiated during the authors' visit to the Nuclear
Theory Institute at the University of Washington and we thank the Institute
for the hospitality extended to us.

\newpage
\begin{center}
{\large{\bf Figure Caption}}
\end{center}
\noindent Figure 1.\ \ \ Comparison of $\Delta W$ (solid line),
the deviation of the approximate and the exact Hermitian effective
interactions, with the upper bound $W_0Z_w$ as a function of the strength
parameter $x$.
\end{document}